\documentclass[aps,pra,epsf,superscriptaddress,amsmath,amssymb,amsfonts,twocolumn,showpacs,floatfix]{revtex4-2}

\usepackage[utf8]{inputenc}

\usepackage{amstext}
\usepackage{amsthm}
\usepackage{physics}

\usepackage{bm} 

\usepackage{graphicx}

\usepackage{epsfig}
\usepackage{dcolumn}
\usepackage{bm}
\usepackage{braket}
\usepackage{color}
\usepackage[colorlinks=true,citecolor=blue]{hyperref}

\usepackage[english]{babel}

\usepackage{multirow}

\usepackage[dvipsnames]{xcolor}

\definecolor{deeppurple}{rgb}{0.7, 0, 0.8}

\allowdisplaybreaks

\usepackage{hyperref}
\hypersetup{      
	urlcolor=blue
}

\usepackage{natbib}

\usepackage{mathtools}
\usepackage[linesnumbered,ruled,vlined]{algorithm2e}
\graphicspath{{./paper_pics/}}
\usepackage{subcaption}

\newcommand{\ct}{\mathcal{T}}
\newcommand{\cv}{\mathcal{V}}
\newcommand{\CL}{\mathcal{L}}

\newcommand{\bc}{\mathbf{c}}

\newcommand{\bp}{\mathbf{p}}
\newcommand{\bq}{\mathbf{q}}
\newcommand{\bx}{\mathbf{x}}

\newcommand{\btheta}{\bm{\theta}}

\newcommand{\R}{\mathbb{R}}

\newcommand{\proj}{\text{\normalfont Proj}}
\newcommand{\Span}{\text{\normalfont span}}

\begin{document}

\title{Machine learning independent conservation laws through neural deflation}

\author{Wei Zhu}
\affiliation{Department of Mathematics and Statistics, University
of Massachusetts Amherst, Amherst, MA 01003-4515, USA}

\author{Hong-Kun Zhang}
\affiliation{Department of Mathematics and Statistics, University
of Massachusetts Amherst, Amherst, MA 01003-4515, USA}

\author{P. G. Kevrekidis}
\affiliation{Department of Mathematics and Statistics, University
of Massachusetts Amherst, Amherst, MA 01003-4515, USA}

\date{\today}

\begin{abstract}
We introduce a methodology for seeking conservation laws
within a Hamiltonian dynamical system,
which we term ``neural deflation''. 
Inspired by deflation methods for
steady states of dynamical systems,
we propose to {iteratively} train a number of
neural networks to minimize a regularized
loss function accounting for the 
necessity of conserved quantities to
be {\it in involution} and 
enforcing functional independence thereof consistently in the infinite-sample limit.
The method is applied to a 
series of integrable and non-integrable
lattice differential-difference equations.
In the former, the predicted number of
conservation laws extensively grows with
the number of degrees of freedom, while
for the latter, it generically stops at
a threshold related to the number of 
conserved quantities in the system.
This data-driven tool could
prove valuable in assessing a model's conserved quantities and its potential integrability.
\end{abstract}

\maketitle

\section{Introduction}
The topic of identification of 
conservation laws and of the potential
integrability of a Hamiltonian dynamical
system has been central to 
 both classical~\cite{Goldstein2001,ablowitz2},
and quantum systems. In particular,
it is expected for a $d$-dimensional
dynamical system that there will 
generically
exist some 
$d/2$ Poisson-commuting (i.e.,
{\it in involution}) conserved
quantities to ensure 
integrability
in the Liouville sense.
Since the relevant settings
arise in a wide variety of 
physical applications including, but
not limited to, optical, atomic,
material, fluid and plasma models~\cite{agrawal,stringari,dauxois,infeld,plasmabook}, such features remain a central  and widely studied topic.

This theme has a time-honored
history and there have been numerous
methods, including ones based
on Painl{e}v{\'e} property~\cite{conte},
as well as ones based on
Lyapunov exponents (see, e.g.,~\cite{benettin:1980a,benettin:1980b}
and~\cite{sandra} for an associated
recent discussion). Nevertheless,
over the past few years, there has been
an extensive effort in this direction
based on the premise of data-driven methods, enabling the identification of
conservation laws via a variety of
machine-learning techniques.
Relevant techniques have extended
from symplectic neural networks
for identifying Hamiltonian dynamical 
systems from data~\cite{karn},
to devising neural transformations
aimed at learning the symmetries
of classical integrable systems~\cite{austen}, and from 
the use of Siamese neural 
networks~\cite{siam}, to optimal 
transport and diffusion maps 
approaches for manifold learning~\cite{marin} (that may work
beyond conservative systems).
They also span the AI Poincar{\'e}
approach learning conservation laws
from trajectories~\cite{teg1} 
and discovering hidden symmetries~\cite{teg2}
to the
most recent and state-of-the-art
approach of learning such conservation
laws from the system's (differential)
equations~\cite{liu2022machine}.
This wide range of efforts indicates
the significance and potential of such
methods, despite possible limitations. Indeed, we are not aware
of methods proposed so far, able to detect 
the progressive increase of conservation
laws, especially  when the number of degrees of freedom
increases. 
We are not familiar with efforts to detect the integrability
of the associated system for a large
number of degrees of freedom. Indeed, when
used for integrable systems, the methods
typically identify a few conservation
laws~\cite{marin,liu2022machine},
which are argued to be relevant
(e.g., physically).

Our aim in the present work is to 
present a method for identifying the
number of conservation laws of
a system, with a view to large(r)
numbers of differential equations.
We are motivated by the notion
of {\it deflation} for steady states
of partial differential equations~\cite{farrell2015deflation},
whereby once a stationary state has been
identified, subsequent iteration steps
weigh against proximity to such
a state, thus discovering additional
ones. Here, we devise a data-driven
methodology in which the regularized 
loss function accounts for two central
features (in our effort to seek
additional conservation laws). Firstly,
these must be in involution with earlier
ones and furthermore, while they are not required to ensure 
{point-wise \textit{orthogonality} in the gradients} (as 
in~\cite{liu2022machine}), they do 
need to enforce {\textit{linear independence} in the gradients}  from 
earlier ones {to achieve \textit{functional independence}}. Iteratively accounting for
these features, we showcase that 
not only can we capture the
appropriate number of conservation laws
for systems previously benchmarked
such as isotropic and anisotropic oscillators
and the three-body problem.
We are also able to do so for 
fundamental differential-difference
integrable and non-integrable systems,
such as the (integrable) 
Toda lattice~\cite{toda}
and (the associated non-integrable, famous)
Fermi-Pasta-Ulam-Tsingou system~\cite{FPUreview} and similarly
for the integrable Calogero 
model~\cite{Calogero71,MOSER1975197},
as well as the discrete sine-Gordon
equation~\cite{dauxois} (of wide
physical relevance to coupled torsion pendula and superconducting Josephson junctions).
\vspace{1em}

\section{Method}
\label{sec:method}
Consider a Hamiltonian system  under the coordinate $\bx \in D\subset \R^{d}$,
\begin{align}
\label{eq:hamiltonian_system}
    d\bx/dt = \mathbf{f}(\bx), \quad \mathbf{f}(\bx)=J(\bx)\nabla H(\bx),
\end{align}
where $H:D\to \R$ is the Hamiltonian function, $J(\bx)\in {\Bbb R}^{d\times
d}$ is an anti-symmetric matrix. 
The Poisson bracket  for two smooth functions
$F$ and $G$ on $D$ takes the form
\begin{equation}\label{poisson}
\{ F,G\} = \nabla F(\bx)^T J(\bx) \nabla G(\bx)
\end{equation}


A \textit{conservation law} is characterized by the vanishing of the Poisson
bracket with $H$. More precisely, a function ${{I}}: D\to\mathbb{R}$ is a conservation law of the
autonomous Hamiltonian system, if 
\begin{equation}\label{eq:conservation_gradient}
\{{{I}}, H\}=\mathbf{f}\cdot \nabla I(\bx)   =0, \,\,\,\forall \bx\in D.\end{equation}
 A collection of $K$ conservation laws $\{I_k\}_{k\in [K]}$, where $[K]\coloneqq \{1, \cdots, K\}$, is said to be \textit{functionally independent} if their gradients $\{\nabla I_k(\bx)\}_{k\in[K]}$ are \textit{linearly independent} as vectors in $\R^d$ for all $\bx\in D$. Furthermore, they are said to be \textit{Poisson-commuting}, or \textit{in involution}, if their pair-wise Poisson bracket vanishes, i.e., for all $j\neq k$,
$ \{I_j, I_k\}
    = 0.
$

Our goal is to use machine learning to obtain a \textit{maximal} set of functionally independent, Poisson-commuting conservation laws $\{I_k(\bx)\}_{k=1}^{d_0}$. If $d_0 = d/2$, then the Hamiltonian system is said to be \textit{integrable} (in the Liouville sense.) It is worth noting that the number $d_0\le d/2$ is generally unknown \textit{a priori}, and the difficulty is to determine $d_0$ in a principled data-driven manner.

\def\comment{
The geometric content of $d_0$-integrability is described by the Liouville-Arnold theorem, which states that 
if the surface of the constant value of the Hamiltonian $H(\bx)=E$
is a  compact connected manifold, then for any fixed array $\bc=(c_1, \cdots, c_{d_0})\in \mathbb{R}^{d_0}$, there is a $d-d_0$ dimensional submanifold (called  isosurface) in $\mathbb{R}^{d}$: 
$$ D_{\bc}: = \{ I_k(\bx) = c_k= {\rm const},\; k=1,\ldots,d_0\}
$$
such that the Hamiltonian system restricted on isosurface $ D_{\bc}$  is ergodic.  In fact, if any point of a
trajectory lies on the isosurface $D_{\bc}$, then all other points
from the trajectory will lie on the same isosurface. Thus if one could identify all the conservation laws, and $d_0>1$, then it is enough to study the dynamical system on  much smaller isosurfaces.
}

A recent attempt to achieve this was proposed by Liu et al.~\cite{liu2022machine}, where they consider  the canonical orthogonality condition \eqref{eq:conservation_gradient}, with a symplectic matrix $J$. Specifically, they randomly sample from the phase space $D\subset \R^d$ a training set $\ct$ and a validation set $\cv$, and parameterize each conserved quantity $I_k(\bx)$ using a neural network $I_k(\bx; \btheta_k)$, where $\btheta_k$ are the trainable network parameters. Since $d_0$ is generally unknown, they simultaneously train a total of $d$ neural networks $\{I_k(\bx; \btheta_k)\}_{k=1}^d$ to minimize the following regularized loss function
\begin{align}
   \nonumber
        \CL(\btheta_1, \cdots, \btheta_d; \ct) &\coloneqq
        \frac{1}{d}\sum_{k=1}^d\ell_{\text{conserv}}[I_k(\cdot; \btheta_k); \ct]
\\
         &+ \lambda \frac{2}{d(d-1)}\sum_{k\neq l}R(\btheta_k, \btheta_l; \ct).
        \label{eq:total_loss_tegmark}
\end{align}
The first term in \eqref{eq:total_loss_tegmark} is the mean of the \textit{conservation losses} for each $I_k$ based on the condition \eqref{eq:conservation_gradient}, i.e.,
\begin{align}
\label{eq:conservation_loss}
    \ell_{\text{conserv}}[I_k(\cdot; \btheta_k); \ct] \coloneqq \frac{1}{|\ct|}\sum_{\bx\in\ct} \left|\widehat{\mathbf{f}}(\bx)\cdot\widehat{\nabla I_k} (\bx; \btheta_k)\right|^2,
\end{align}
where $\widehat{\mathbf{f}}(\bx)$ and $\widehat{\nabla I_k}(\bx; \btheta_k)$ are, respectively, the $l^2$-normalized vectors of $\mathbf{f}(\bx)$ and $\nabla I_k(\bx; \btheta_k)$. The second term in \eqref{eq:total_loss_tegmark} is a regularization to encourage functional independence among $\{I_k(\cdot; \btheta_k)\}_{k=1}^d$ by enforcing point-wise \textit{orthogonality} among the gradients $\{\nabla I_k(\cdot; \btheta_k)\}_{k=1}^d$,
\begin{align}
\label{eq:independent_loss}
    R(\btheta_k, \btheta_l; \ct) \coloneqq \frac{1}{|\ct|}\sum_{\bx\in\ct}\left| \widehat{\nabla I_{k}}(\bx; \btheta_k)\cdot\widehat{\nabla I_l} (\bx; \btheta_l)\right|^2.
\end{align}
Once $\{I_k(\bx; \btheta_k^*)\}_{k=1}^d$ are trained, they consider the following Jacobian matrices on the validation set $\cv$.
\begin{align}
    \left[\nabla I_1(\bx; \btheta_1^*), \cdots, \nabla I_d(\bx; \btheta_d^*)\right]\in \R^{d\times d}, \quad \bx\in \cv,
\end{align}
and a maximal functionally independent subset of the learned $\{I_k(\bx; \btheta_k^*)\}_{k=1}^d$ is obtained by identifying the largest set of columns $\mathcal{K}\subset \{1, \cdots, d\}$ of the above matrices that are consistently linearly independent for all $\bx\in \cv$. The cardinality $|\mathcal{K}|$ is then declared as the number $d_0$ of the independent conservation laws of system.

\begin{figure*}[tbp!]
  \centering
  \begin{subfigure}[b]{0.32\textwidth}
    \centering
    \includegraphics[width=\textwidth]{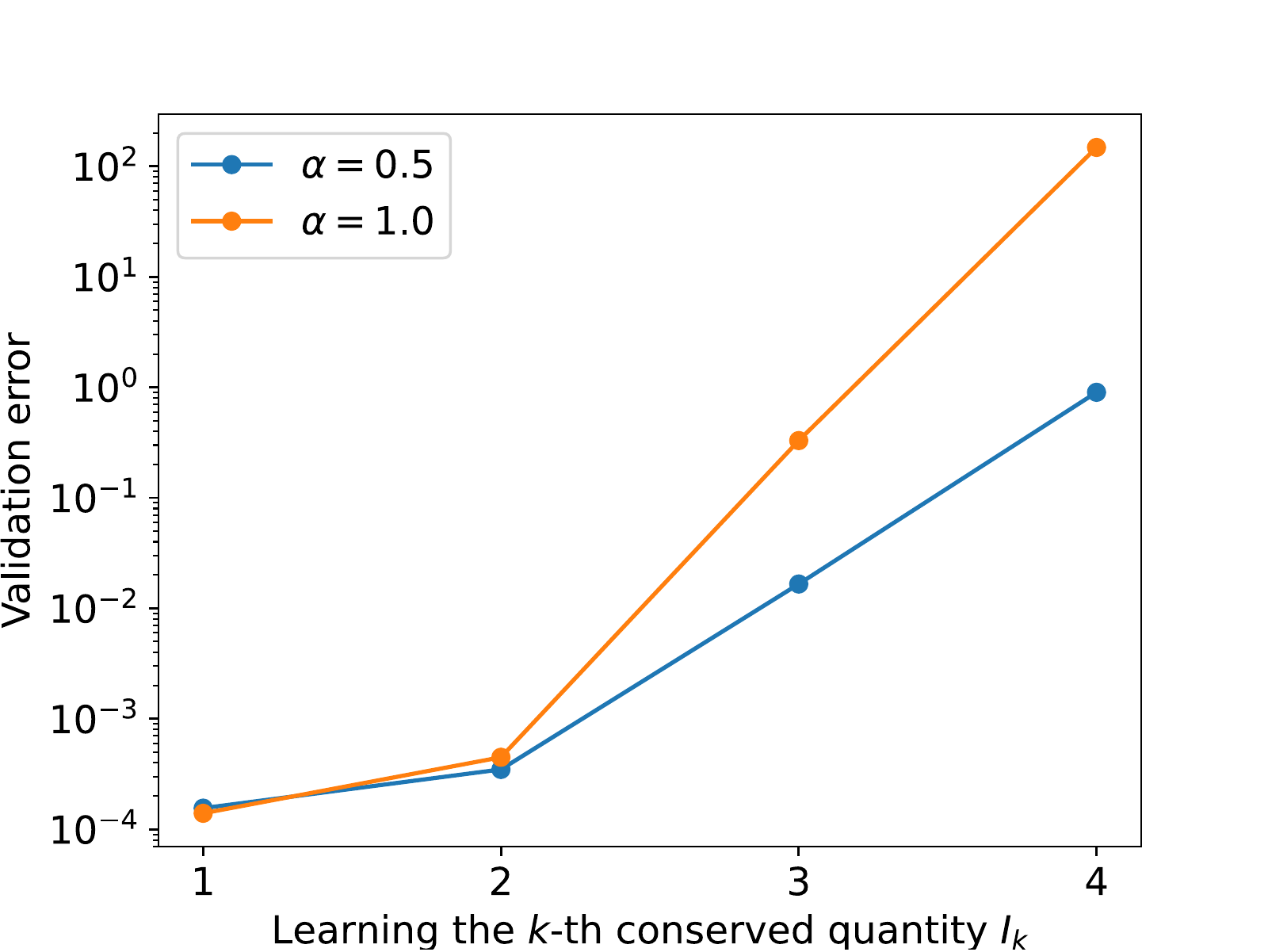}
    \caption{Isotropic harmonic oscillator.}
    \label{fig:iso_osci}
  \end{subfigure}
    \begin{subfigure}[b]{0.32\textwidth}
    \centering
    \includegraphics[width=\textwidth]{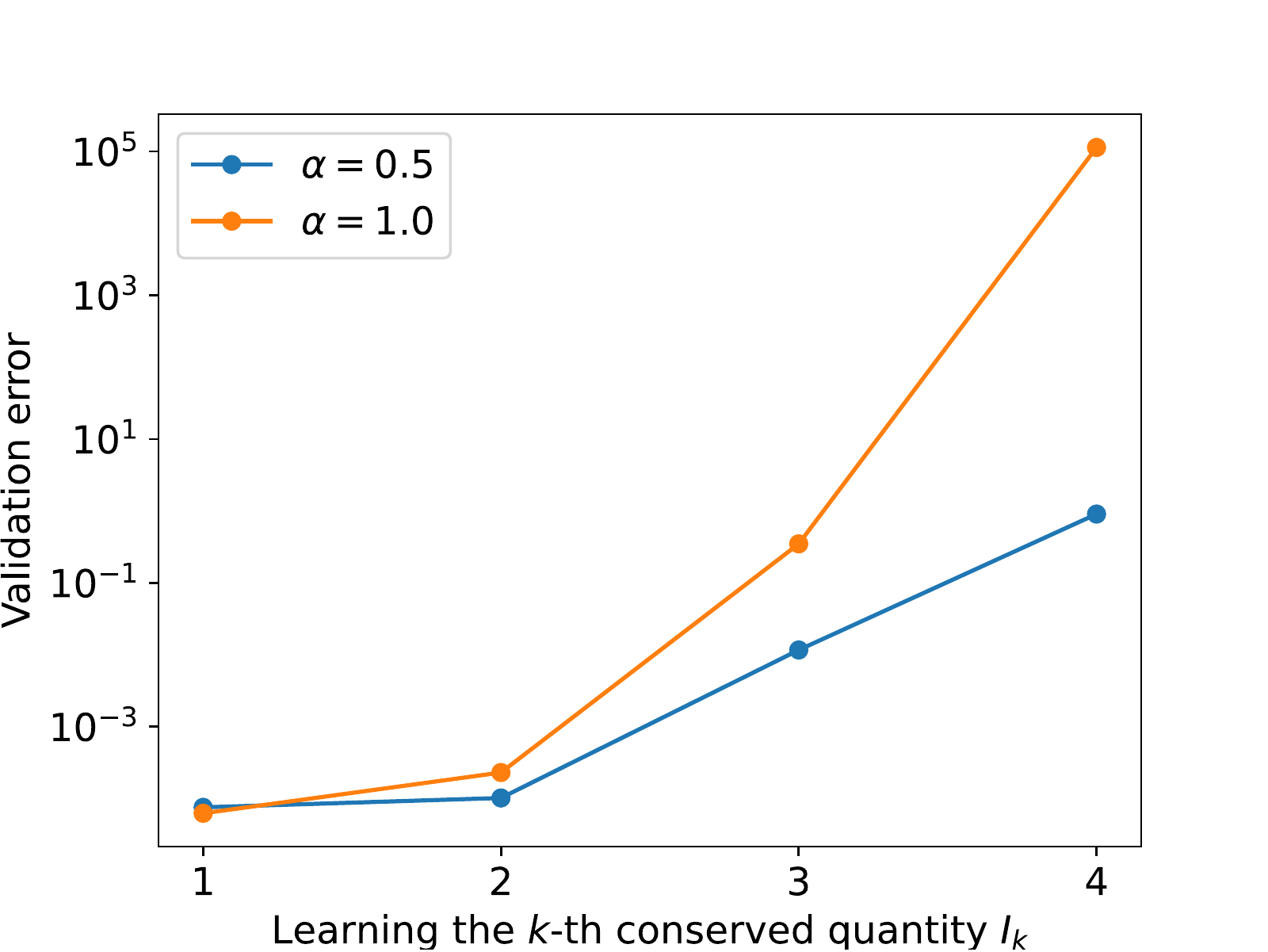}
    \caption{Anisotropic harmonic oscillator.}
    \label{fig:aniso_osci}
  \end{subfigure}
    \begin{subfigure}[b]{0.32\textwidth}
    \centering
    \includegraphics[width=\textwidth]{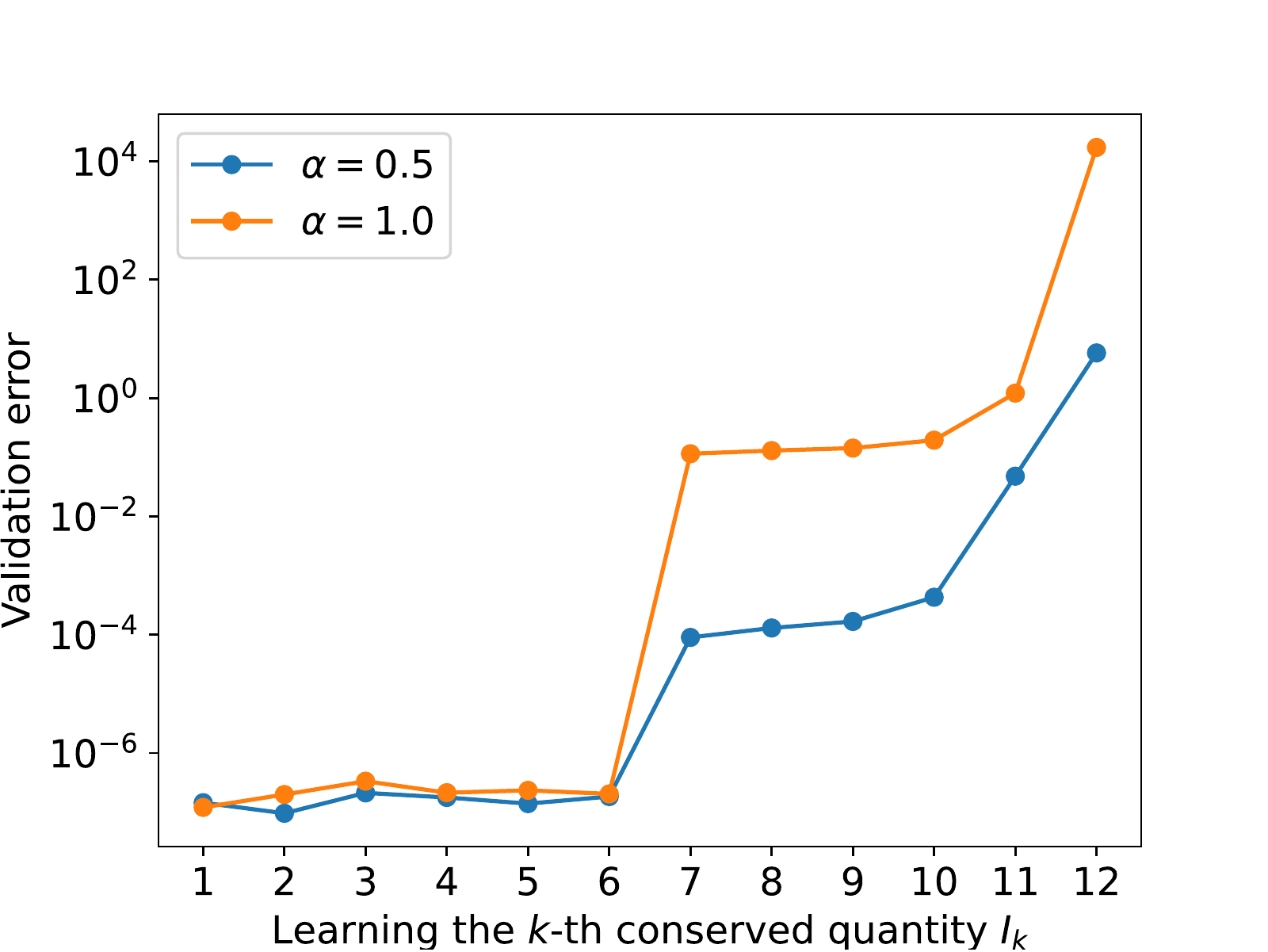}
    \caption{2D Three-body problem.}
    \label{fig:3body}
  \end{subfigure}
  \caption{The validation losses $\{\CL_k(\btheta_k^*; \cv)\}_{k=1}^d$ of the conserved quantities $\{I_k(\cdot; \btheta_k^*)\}_{k=1}^d$ trained by Algorithm~\ref{alg:deflation} for the three integrable Hamiltonian systems, where $d$ is the number of degrees of freedom.  A significant jump in loss (of several orders of magnitude) occurs at $k=d/2+1$ in each case. This jump indicates that our method has accurately predicted the integrability of the systems. This result is consistent across different choices of deflation strength, namely $\alpha=1.0$ or $\alpha=0.5$, although the increase in validation loss is more pronounced with larger deflation strength.}
  \label{fig:results_group_1}
\end{figure*}

Although the above method offers a rough estimation of the independent conservation laws $\{I_k\}_{k=1}^{d_0}$ and their total number $d_0$, it suffers from several limitations. Firstly, the regularization term in \eqref{eq:total_loss_tegmark} encourages point-wise \textit{orthogonality} among the gradients $\{\nabla I_k(\bx; \btheta_k)\}_{k=1}^d$. However, it is important to note that functional independence of $\{I_k(\cdot; \btheta_k)\}_{k=1}^d$ merely implies \textit{linearly independent} gradients, which are generally \textit{not} orthogonal.  Therefore, the loss function \eqref{eq:total_loss_tegmark} is \textit{inconsistent} in the sense that it does not vanish even if we plug into \eqref{eq:total_loss_tegmark} the \textit{ground-truth} set of $d_0$ independent conservation laws $\{I_k(\bx)\}_{k=1}^{d_0}$. As a result, one cannot declare with high confidence that the trained networks are indeed independent conservation laws simply by examining the magnitude of the loss \eqref{eq:total_loss_tegmark} on the validation set. Secondly, Eq.~\eqref{eq:total_loss_tegmark} does not require $\{I_k(\cdot; \btheta_k)\}_{k=1}^d$ to be \textit{in involution}. Consequently, the number of independent conservation laws detected in \cite{liu2022machine} is sometimes much larger than $d/2$, whereas there should be \textit{at most} $d/2$ in involution.

\vspace{.2em}
{\bf Neural deflation method.} 
In light of the above issues, we propose the  \textit{neural deflation method} to iteratively learn each $I_k(\cdot; \btheta_k)$ in a principled and interpretable manner, for general Hamiltonian system (\ref{eq:hamiltonian_system}). The benefit of our method is that the loss function for each $I_k(\cdot; \btheta_k)$ is close to zero if only if there exist at least $k$ independent conservation laws in involution. Therefore, the number of the conservation laws can be determined by identifying the index $k\in\{1, \cdots, d\}$ after which there is a significant jump in the validation loss.

Specifically, to learn the first conservation law $I_1(\cdot; \btheta_1)$, we minimize the following loss function $\CL_1(\btheta_1; \ct)$ based solely on the orthogonality condition \eqref{eq:conservation_gradient},
\begin{align}
\label{eq:l1}
    \CL_1(\btheta_1; \ct)& \coloneqq \ell_\text{conserv}[I_1(\cdot; \btheta_1); \ct],
\end{align}
where $\ell_{\text{conserv}}$ is defined in \eqref{eq:conservation_loss}, for general anti-symmetric matrix $J$.
In fact, since the Hamiltonian $H$ is always a conserved quantity for \eqref{eq:hamiltonian_system}, we can in principle directly set  $I_1(\bx)$ to be $H(\bx)$ without parameterizing it as a neural network. However, training $I_1(\bx; \btheta_1)$ based on \eqref{eq:l1} can provide us a gauge on the magnitude of the training/validation loss in order to learn and identify the subsequent conserved quantities.

We then inductively learn a sequence of conservation laws as follows. Assuming we have already obtained $K-1$ conserved quantities $\{I_k(\bx;\btheta_k^*)\}_{k=1}^{K-1}$, where $K\ge 2$, we train the $K$-th conservation law $I_K(\bx;\btheta_{K})$ using the following \textit{deflated} loss function $\CL_K(\btheta_K; \ct)$  while fixing the learned parameters $\{\btheta_k^*\}_{k=1}^{K-1}$ of the previous networks,
\begin{align}
    \label{eq:lK}
    & \CL_{K}(\btheta_K; \ct)\\
    \nonumber
    \coloneqq  & \frac{ \left(\ell_{\text{conserv}}[I_K(\cdot; \btheta_K); \ct] + \sum_{k=1}^{K-1}\ell_{\text{inv}}[\btheta_k^*, \btheta_K; \ct]\right)/K}{\displaystyle\left|\frac{1}{|\ct|}\sum_{\bx\in \ct} \left\|\proj_{\Span\left\{\widehat{\nabla I_k}(\bx; \btheta_k^*)\right\}_{k\in [K-1]}^\perp}\widehat{\nabla I_K} (\bx; \btheta_K)\right\|^2\right|^\alpha},
\end{align}
where
\begin{align*}    
    \ell_{\text{inv}}[\btheta_k^*, \btheta_K; \ct]
    =\frac{1}{|\ct|}\sum_{\bx\in\ct}\left|\left\{I_1(\cdot; \btheta_k^*), I_2(\cdot; \btheta_K)\right\}(\bx)\right|^2
\end{align*}
ensures that $I_K(\cdot; \btheta_K)$ Poisson-commutes with all previously learned $\{I_k(\cdot; \btheta_k^*
)\}_{k=1}^{K-1}$, and the loss is divided by $K$ due to having $K$ terms in total in the numerator. The denominator is a deflation factor that enforces \textit{functional independence} between $I_K(\cdot; \btheta_K)$ and $\{I_k(\cdot; \btheta_k^*
)\}_{k=1}^{K-1}$; more specifically, the operator
\begin{align}
\label{eq:deflation_denominator}
    \proj_{\Span\left\{\widehat{\nabla I_k}(\bx; \btheta_k^*)\right\}_{k\in[K-1]}^\perp}\widehat{\nabla I_K} (\bx; \btheta_K)
\end{align}
denotes the projection of the vector $\widehat{\nabla I_K} (\bx; \btheta_K)$ onto the orthogonal complement of the subspace in $\R^d$ spanned by $\{\widehat{\nabla I_k} (\bx; \btheta_k^*)\}_{k\in [K-1]}$. We note that for \eqref{eq:deflation_denominator} to be nonzero, in contrast to Eq.~\eqref{eq:independent_loss}, $\widehat{\nabla I_K} (\bx; \btheta_K)$ is only required to be linearly independent with the previously learned $\{\widehat{\nabla I_k} (\bx; \btheta_k^*)\}_{k\in [K-1]}$ instead of being orthogonal. Finally, the deflation power $\alpha>0$ is a hyperparameter adjusting the strength of the constraint on functional independence between $I_K(\cdot; \btheta_K)$ and $\{I_k(\cdot; \btheta_k^*)\}_{k\in [K-1]}$.

Compared to the model \eqref{eq:total_loss_tegmark} in \cite{liu2022machine}, our model \eqref{eq:lK} has the clear advantage of being \textit{consistent} in the infinite-sample limit. More specifically, assuming that the previously obtained $\{I_k(\cdot;\btheta_k^*)\}_{k=1}^{K-1}$ perfectly parameterize a ground-truth set of independent conservation laws in involution and that the empirical sums $\frac{1}{|\ct|}\sum_{\bx\in\ct}[\cdots]$ are replaced by the expectations $\mathbb{E}_{\bx\sim \mu}[\cdots]$ for some absolutely continuous probability measure $\mu$ over the phase space $D\subset\R^d$, then $I_K(\cdot; \btheta_K)$ achieves a zero loss in the infinite limit \textit{if and only if} $\{I_k(\cdot;\btheta_k)\}_{k=1}^{K}$ is a set of $K$ independent Poisson-commuting conservation laws.

We repeat the process until we observe a significant increase in the loss function $\CL_K(\btheta_K^*; \cv)$ on the validation set $\cv$, and declare, at this point, $\{I_k(\cdot; \btheta_k^*)\}_{k=1}^{K-1}$ as a maximal set of $d_0=K-1$ independent Poisson-commuting conservation laws of the system. Our method is summarized in Algorithm~\ref{alg:deflation}.

\begin{figure*}[tbp!]
  \centering
  \begin{subfigure}[b]{0.32\textwidth}
    \centering
    \includegraphics[width=\textwidth]{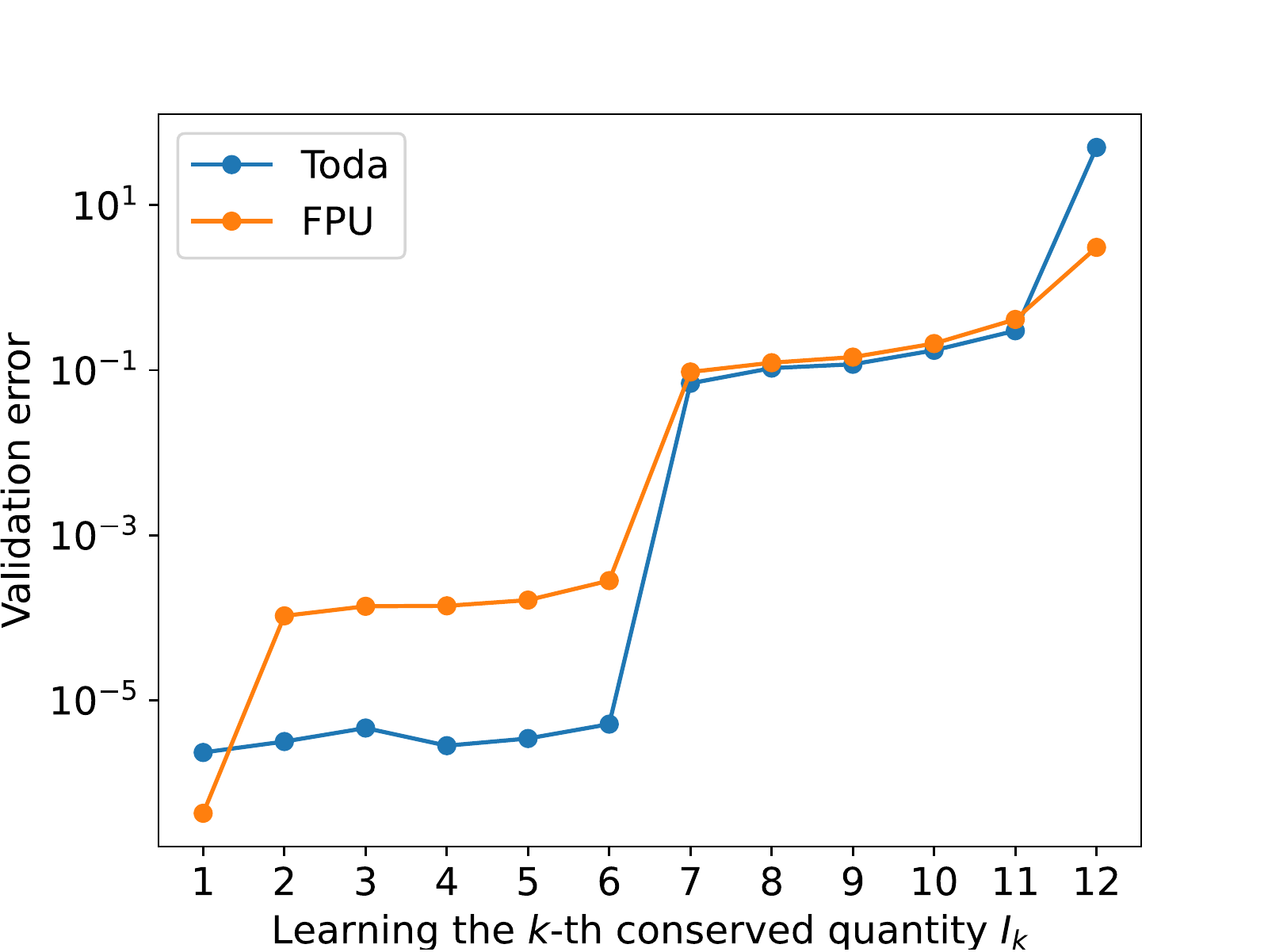}
    \caption{Number of lattice sites = 6.}
    \label{fig:toda_fput_6}
  \end{subfigure}
  \begin{subfigure}[b]{0.32\textwidth}
    \centering
    \includegraphics[width=\textwidth]{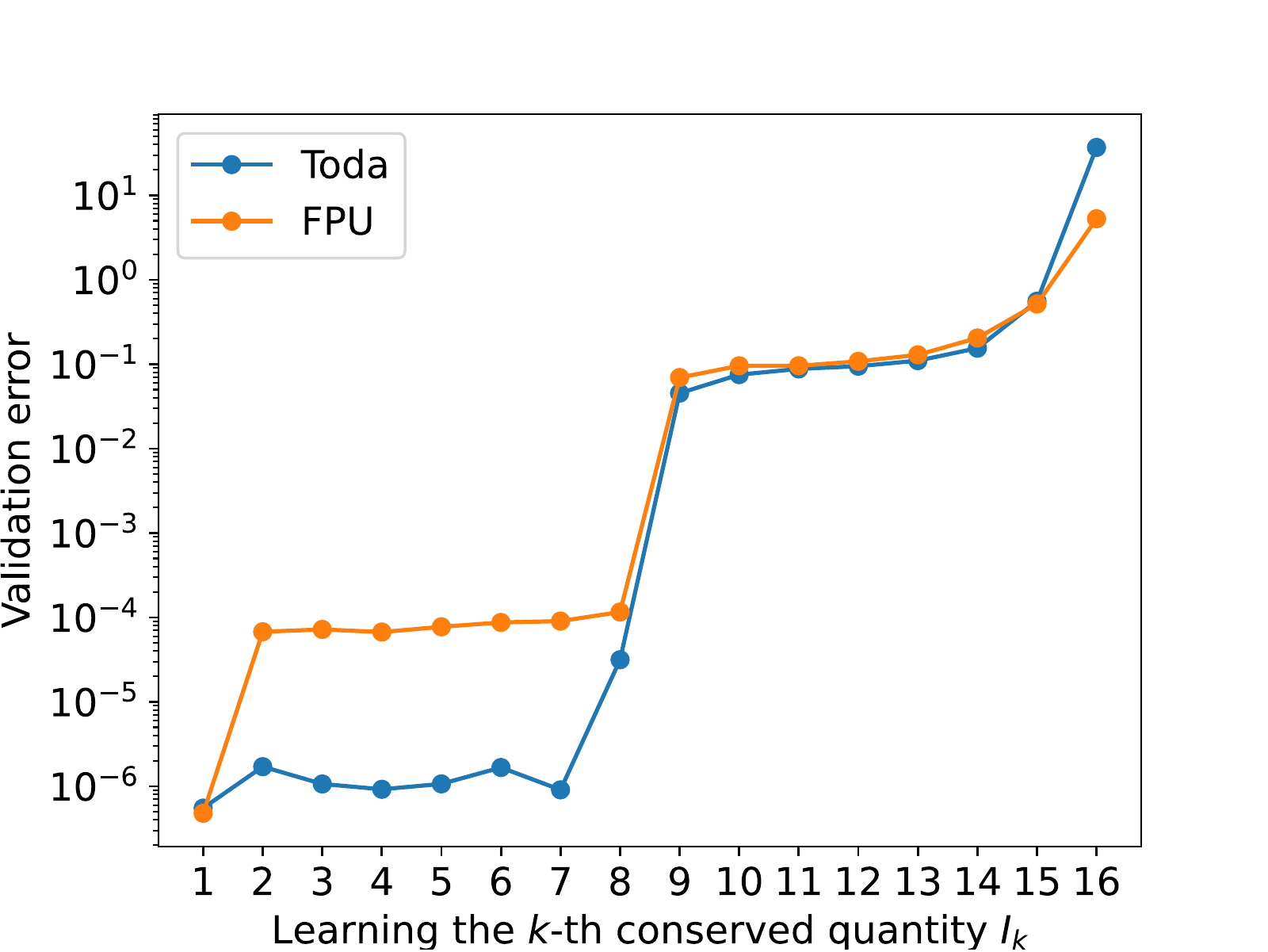}
    \caption{Number of lattice sites = 8.}
    \label{fig:toda_fput_8}
  \end{subfigure}
  \begin{subfigure}[b]{0.32\textwidth}
    \centering
    \includegraphics[width=\textwidth]{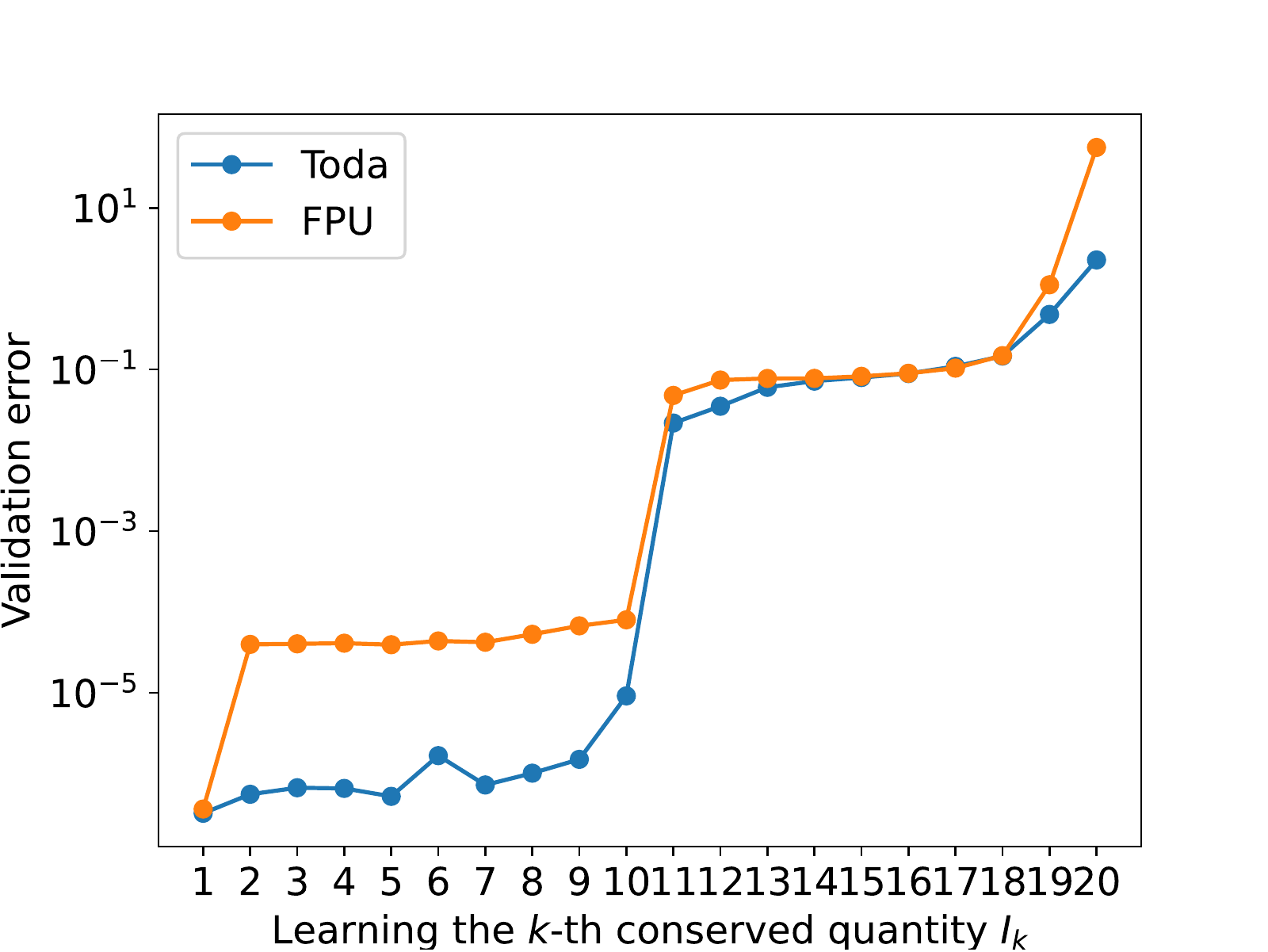}
    \caption{Number of lattice sites = 10.}
    \label{fig:toda_fput_10}
  \end{subfigure}  
  \caption{The validation losses $\{\CL_k(\btheta_k^*; \cv)\}_{k=1}^d$ of the conserved quantities $\{I_k(\cdot; \btheta_k^*)\}_{k=1}^d$ trained by Algorithm~\ref{alg:deflation} for the integrable Toda lattice and the non-integrable FPUT system with varying degrees of freedoms $d=2N$, where $N$ is the number of lattice sites.  For the Toda system, the validation loss consistently exhibits a significant jump at $k=d/2+1$ for varying degrees of freedom $d$, whereas the jump occurs at $k=2$ for the FPUT systems. See Section~\ref{sec:experiments} for a detailed explanation of the results.}
  \label{fig:results_toda_fput}
\end{figure*}

\begin{figure*}[tbp!]
  \centering
    \begin{subfigure}[b]{0.32\textwidth}
    \centering
    \includegraphics[width=\textwidth]{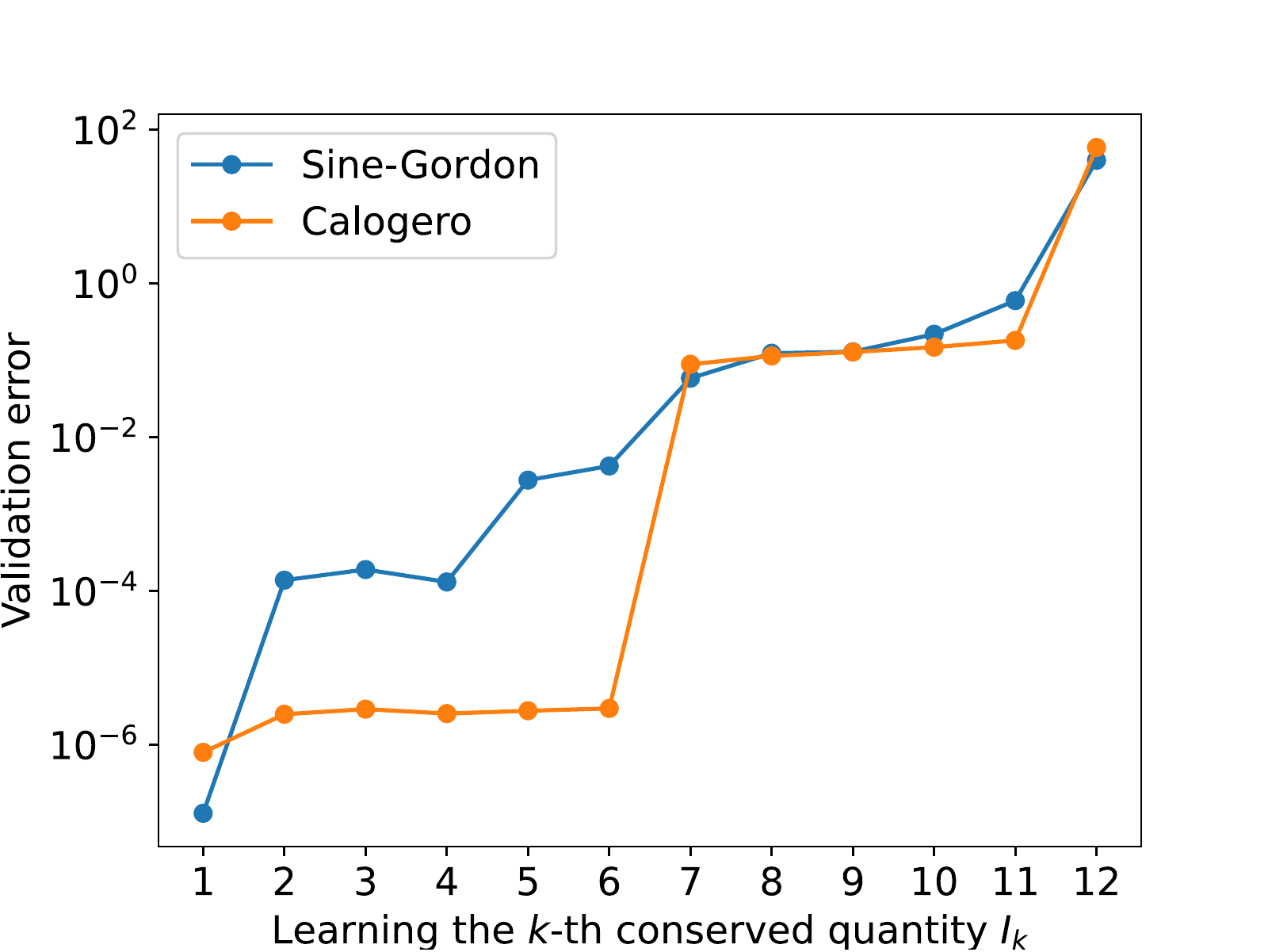}
    \caption{Number of lattice sites = 6.}
    \label{fig:sg_calogero_6}
  \end{subfigure}
  \begin{subfigure}[b]{0.32\textwidth}
    \centering
    \includegraphics[width=\textwidth]{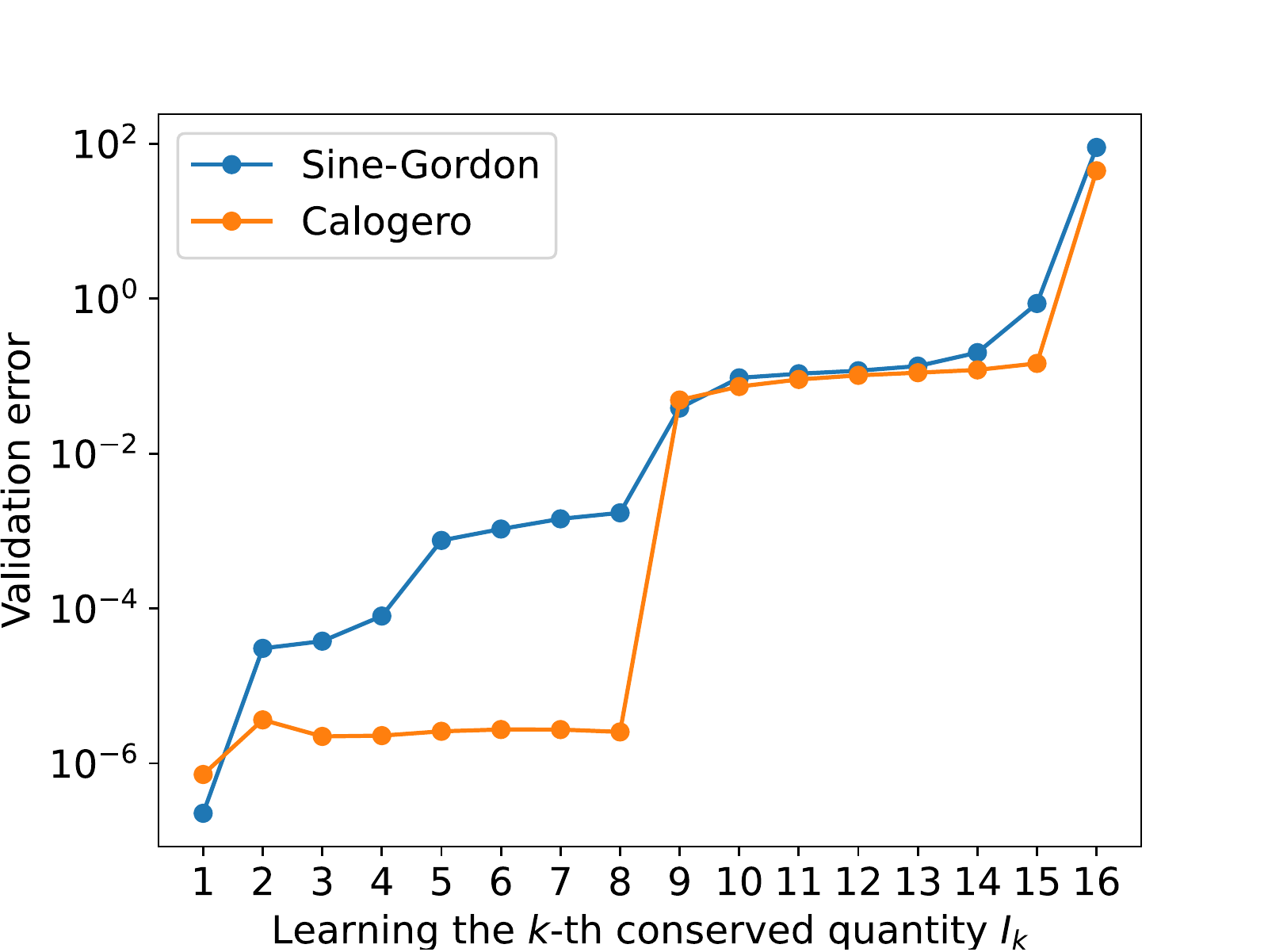}
    \caption{Number of lattice sites = 8.}
    \label{fig:sg_calogero_8}
  \end{subfigure}
    \begin{subfigure}[b]{0.32\textwidth}
    \centering
    \includegraphics[width=\textwidth]{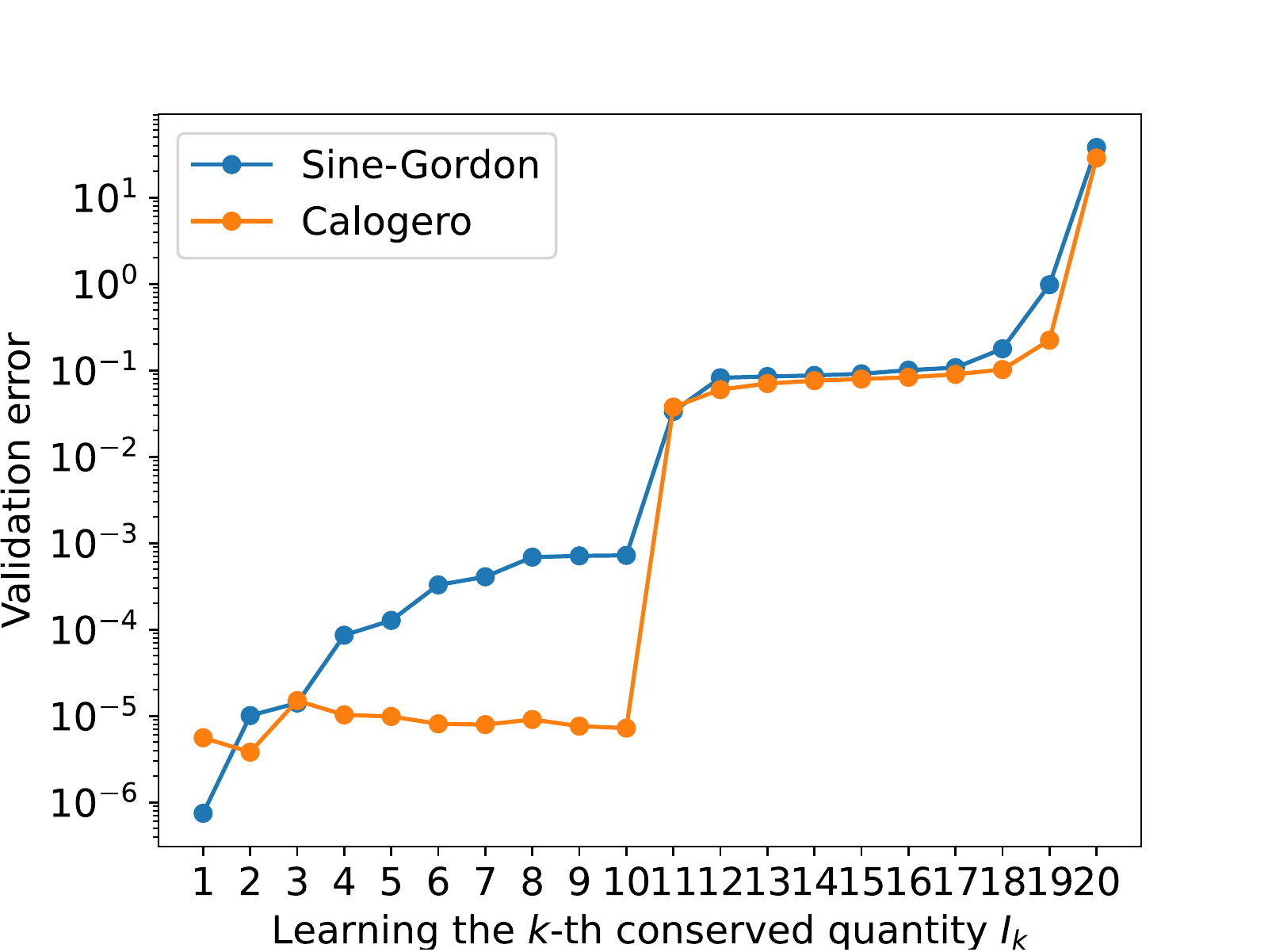}
    \caption{Number of lattice sites = 10.}
    \label{fig:sg_calogero_10}
  \end{subfigure}
  \caption{The validation losses $\{\CL_k(\btheta_k^*; \cv)\}_{k=1}^d$ of the conserved quantities $\{I_k(\cdot; \btheta_k^*)\}_{k=1}^d$ trained by Algorithm~\ref{alg:deflation} for the non-integrable discrete sine-Gordon system and the integrable Calogero's system with varying degrees of freedoms $d=2N$, where $N$ is the number of lattice sites.  For the Calogero's system, the validation loss consistently exhibits a significant jump at $k=d/2+1$ for varying degrees of freedom $d$, whereas the jump occurs at $k=2$ for the sine-Gordon systems. See Section~\ref{sec:experiments} for a detailed explanation of the results.}
  \label{fig:results_sg_calogero}
\end{figure*}

\begin{algorithm}[h]
  \SetAlgoLined
  \KwIn{Hamiltonian system $d\bx/dt = \mathbf{f}(\bq, \bp)$, where $\bx = (\bq, \bp)\in D\subset \R^d$.}
  \KwOut{A maximal set $\{I_k(\cdot; \btheta_k^*)\}_{k=1}^{d_0}$ of independent conservation laws in involution. }
  \BlankLine
  Randomly sample a training set $\ct$ and a validation set $\cv$  from the phase space $D\subset \R^d$\;
  $\btheta_1^* \leftarrow \arg\min_{\btheta_1}\CL_1(\btheta_1; \ct)$ given by Eq.~\eqref{eq:l1}\;
  $\CL_1^\text{val} \leftarrow \CL_1(\btheta_1^*; \cv)$\;
  $K\leftarrow 1$\;
  \Repeat{$\CL_K^{\text{\normalfont val}}/\CL_1^{\text{\normalfont val}} > \text{\normalfont tol}$}{
    $K\leftarrow K+1$\;
    $\btheta_K^* \leftarrow \arg\min_{\btheta_K}\CL_K(\btheta_K; \ct)$ given by \eqref{eq:lK}\;
    $\CL_K^\text{val} \leftarrow \CL_K(\btheta_K^*; \cv)$\;
  }
  $d_0\leftarrow K-1$\;
  \caption{Neural deflation method}
  \label{alg:deflation}
\end{algorithm}

\section{Numerical experiments}
\label{sec:experiments}

We present the results of our algorithm in learning independent conservation laws of the 2D isotropic and anisotropic harmonic oscillators, the three-body problem, the Toda and the Fermi-Pasta-Ulam (FPUT) lattices, the discrete sine-Gordon system, and the Calogero's problem.

\vspace{.2em}
{\bf The 2D isotropic/anisotropic oscillators and the three-body examples.}
All three systems are fully integrable in $\R^d$, where $d=4$ and $12$ for the harmonic oscillators and the three-body systems, respectively~\cite{liu2022machine}. However, we pretend to be agnostic about their integrability,  and use Algorithm~\ref{alg:deflation} to obtain a maximal set of functionally independent Poisson-commuting conservation laws.

We use a 4-layer feedforward neural network with Sigmoid Linear Unit (SiLU) activations and 400 neurons per layer to parameterize each conserved quantity. To train each network, we use the ADAM optimizer \cite{kingma20153rd} for 10,000 iterations with a batch size of 500. We randomly sample 200,000 phase points from the cell $[-1000, 1000]^d$ and divide them equally between the training set $\ct$ and the validation set $\cv$. We compare the results of setting the deflation power $\alpha$ in \eqref{eq:lK} to either $1.0$ or $0.5$.

Figure~\ref{fig:results_group_1} displays the validation losses $\{\CL_k(\btheta_k^*; \cv)\}_{k=1}^{d}$ of the learned conserved quantities $\{I_k(\cdot; \btheta_k^*)\}_{k=1}^d$ trained using Algorithm~\ref{alg:deflation}. For each system, a substantial increase (of several orders of magnitude) in the validation loss occurs precisely at $k = d/2+1$, which indicates that our algorithm has accurately predicted the integrability of the systems (cf.~the last line of Algorithm~\ref{alg:deflation}), and successfully learned a maximal set of independent conservation laws in involution. The numerical results are consistent across different choices of deflation strength, namely $\alpha = 1.0$ or $\alpha=0.5$, although a larger $\alpha$ leads to a more significant increase in validation loss at $k = d/2+1$.

\vspace{.2em}
{\bf The Toda lattice and the FPUT system.} 
We consider the integrable Toda lattice and the associated non-integrable FPUT system with different degrees of freedom $d=2N$, where $N$ is the number of the lattice sites ranging from $5$ to $10$ and periodic boundary conditions. We use a similar experimental setup, but sample the phase points from $[-50, 50]^d$. Deflation strength was only set to $\alpha=1.0$ based on the previous experiment, and the results are shown in Figure~\ref{fig:results_toda_fput}. For the Toda lattice, the validation loss again significantly increases at $k= d/2+1$, (although the jump is not as ``sharp", e.g., for panel (b) 
at $k=d/2$). This implies once again that our method accurately predicts the integrability of the Toda system and learns all the independent conservation laws. Conversely, for the FPUT system, the validation loss consistently jumps at $k=2$ for varying $d$. This means our algorithm accurately predicts the non-integrability of the system and that the number of independent conservation laws remains constant across different degrees of freedom $d$. However, we note that the FPUT system actually has $d_0=2$ independent conservation laws (i.e., the momentum and $H$) instead of the predicted $(k-1)=1$ learned by our algorithm. Nonetheless, the distinct behavior of the loss functions between the two systems with varying degrees of freedom highlights the potential of our algorithm in evaluating a system's integrability.

{\bf The discrete sine-Gordon system and Calogero's problem.} Finally, we apply our algorithm to the non-integrable discrete sine-Gordon system and the integrable Calogero's problem with varying degrees of freedom $d$. Even though these two systems are not related, we plot the results in the same figures (Figure~\ref{fig:results_sg_calogero}) to highlight the distinct behavior of the validation losses for an integrable vs. a
non-integrable system. Similar to the previous experiment, for the (integrable) Calogero's problem, the validation loss consistently exhibits a substantial increase at $k=d/2+1$ with varying degrees of freedom $d$. In contrast, for the (non-integrable) sine-Gordon system, the loss always jumps at $k=2$, which is consistent with the fact that
the underlying system only has \textit{one} independent conservation law ($H$), regardless of
the lattice size.

\section{Conclusions \& Future Challenges}
In the present work we have revisited
the extensively studied in recent years
topic of identifying conservation laws
and, ultimately, gauging the potential
integrability of a Hamiltonian model. 
The main contribution of the present
work lies in the introduction of the
technique of neural deflation. Motivated
by recent developments in numerical 
bifurcation analysis, we propose a
technique whose regularized loss 
function involves the involution required
of the integrals of the motion and the
imposition, motivated by deflation,
of their linear independence. We have
shown that the technique works in 
``standard'', previously used examples
such as the isotropic and anisotropic
harmonic oscillator and the three-body problem~\cite{liu2022machine}.
Importantly, though, it successfully 
enables the  consideration of 
higher-dimensional lattice nonlinear
dynamical systems of both integrable
(Toda, Calogero) and nonintegrable
(FPUT, discrete sine-Gordon) type. 
In all the systems examined, we 
saw a distinctive increase (jump) of the
loss function in the vicinity of the
expected number ($d_0$ or just ``a couple'')
of independent conservation laws. 

Admittedly, this direction of research
warrants further efforts. Examining
numerous additional examples, including
continuum ones, will be informative 
towards features such as the ``sharpness''
of the jump and the potential issues 
with capturing all the associated
conservation laws (cf. the FPUT example).
Another important direction is that
of associating the identified 
quantities (and the hypersurfaces they
represent) via symbolic regression
to the actual conserved quantities 
known physically, or identified via
integrability techniques in the systems
of interest. Studies along this vein
are currently in progress and will be
reported in future work.

\bibliography{mybib}

\end{document}